\documentclass[a4paper,twocolumn,english,aps,pre,a4paper,showpacs,floatfix]{revtex4}
\usepackage[T1]{fontenc}
\usepackage[latin1]{inputenc}
\usepackage{graphicx}
\usepackage{epsfig}
\usepackage{bm}
\usepackage{amsmath}
\usepackage{amssymb}

\begin{document}

\title{Biaxial nematic phase stability and demixing behaviour 
in monolayers of rod-plate mixtures}

\author{Yuri Mart\'{\i}nez-Rat\'on}
\email{yuri@math.uc3m.es}
\affiliation{Grupo Interdisciplinar de Sistemas Complejos (GISC), Departamento de Matem\'aticas, Escuela Polit\'ecnica Superior,
Universidad Carlos III de Madrid, Avenida de la Universidad 30, E-28911, Legan\'es, Madrid, Spain}

\author{Miguel Gonz\'alez-Pinto}
\email{miguel.gonzalezp@uam.es}
\affiliation{Departamento de F\'{\i}sica Te\'orica de la Materia Condensada,
Universidad Aut\'onoma de Madrid, E-28049 Madrid, Spain}

\author{Enrique Velasco}
\email{enrique.velasco@uam.es}
\affiliation{Departamento de F\'{\i}sica Te\'orica de la Materia Condensada, 
Instituto de Ciencia de Materiales Nicol\'as
Cabrera and Condensed Matter Physics Center (IFIMAC), Universidad Aut\'onoma de Madrid, E-28049 Madrid, Spain}

\begin{abstract}
We theoretically study the phase behaviour of monolayers of hard rod-plate mixtures using a fundamental-measure 
density functional in the restricted-orientation (Zwanzig) approximation. Particles can rotate in 3D 
but their centres of mass are constrained to be on a flat surface. In addition, we consider both species to be
subject to an attractive potential proportional to the particle contact area on the surface and 
with adsorption strengths that depend on the species type. Particles have board-like shape, with sizes chosen using a symmetry 
criterion: same volume and same aspect ratio $\kappa$. Phase diagrams were calculated
for $\kappa=10$, 20 and 40 and different values of adsorption strengths. For small 
adsorption strengths the mixtures exhibit a second-order uniaxial nematic-biaxial nematic
transition for molar fraction of rods $0\leq x\lesssim 0.9$. In the uniaxial nematic phase 
the particle axes of rods and plates are aligned perpendicular and parallel to the monolayer, 
respectively.  At the transition, the orientational symmetry of the plate axes is broken, and they orient parallel 
to a director lying on the surface. For large and equal adsorption strengths the mixture demixes at low pressures into a uniaxial 
nematic phase, rich in plates, and a biaxial nematic phase, rich in rods. The demixing transition is located 
between two tricritical points. Also, at higher pressures and in the plate-rich part of the phase diagram, the 
system exhibits a strong first-order uniaxial nematic-biaxial nematic phase transition with a large density 
coexistence gap. When rod adsorption is considerably large while that of plates is small, the transition to the 
biaxial nematic phase is always of second order, and its region of stability in the phase diagram
considerably widens. At very high pressures the mixture can effectively be identified as a two-dimensional 
mixture of squares and rectangles which again demixes above a certain critical point. We also studied the 
relative stability of uniform phases with respect to density modulations of smectic, columnar and crystalline 
symmetry.
\end{abstract}

\pacs{61.30.Pq,64.70.M-,47.57.J-}

\maketitle

\section{Introduction}
\label{intro}

It is well known that the properties of biological vesicles and membrane cells strongly depend on the 
their constituent blocks, usually composed of phospholipid bilayers with embedded proteins. 
These molecules are in general anisotropic (rod or plate-like) and the demixed states usually have 
liquid-crystal symmetries, such as isotropic (I), nematic (N) or biaxial nematic (B) symmetries. 
For certain conditions these complex mixtures of biomolecules phase separate, creating regions rich 
in different species and consequently changing the membrane curvature. There is much experimental evidence 
of demixing transitions in monolayers and bilayers of mixed anisotropic biomolecules  
\cite{Sharma,Rice,Hagen,Keller1,Keller2,Baukina,Zanghellini}. 
The adsorption of a large variety of  mixtures of rod-like molecules in Langmuir monolayers has been 
extensively studied both experimentally and theoretically. Many of these works focus on the chemical and 
thermodynamic conditions for which the monolayers become spatially heterogeneous, i.e. when the mixture 
demixes in different phases usually possessing liquid-crystal ordering \cite{Ries,Smith,Martynski,Muller}. 
Finally many experiments on the adsorption of rod-like colloidal particles at the interfaces separating two 
immiscible fluids  showed the propensity of these particles to self-assemble into clusters of different 
geometries \cite{Kim1,Kalashnikova,Kim2}. The degree of adsorption of these particles at the interface and 
their relative orientation with respect to it strongly depend on their chemical compositions. This in turn 
can modify their wetting properties and consequently the effective capillary forces acting between particles. 
The resulting effect is the existence of anisotropy in the pair interaction potential which forces the 
particles to self-assemble into clusters \cite{Kim1,Vora,Davies}.  
When colloids with very different chemical properties are adsorbed at the interface they usually phase separate 
into phases with different composition of species. A recent experimental work showed how demixing of adsorbed 
colloids strongly modifies their self-assembling properties \cite{Choudhuri}.

All the systems discussed above share the following properties: (i) they are mixtures of  
anisotropic particles, (ii) the degrees of freedom of their centres of mass are strongly restricted,  
usually resulting in an effective two-dimensional fluid, and (iii) the particle axes can rotate in 3D but 
with certain restrictions which depend on the degree of particle adsorption on the monolayer, bilayer or 
interface. The main motivation of the present work is the formulation of a very simple model for a binary 
mixture of anisotropic particles (specifically a mixture of rods and plates) which allows a detailed study of the 
conditions (particle aspect ratios, degree of adsorption) under which these mixtures demix into two 
different phases. 

To this purpose we choose particles to have a board-like shape and interact through a hard-core repulsion. Also, for 
simplicity, we use the Zwanzig approximation to account for the orientational degrees of freedom, which are restricted to be
three. Finally, we will use a mean-field density functional (DF) based on the fundamental-measure theory (FMT), derived for 
the present model in the late 90's \cite{cuesta1} and more recently implemented to 
calculate the phase diagrams of rod and plate board-like particles \cite{yuri0}. Monolayers 
of one-component rods or plates were recently studied within this theory, considering uniaxial 
\cite{yuri1} and biaxial \cite{miguel1} particle geometries. In the latter work, phase diagrams were calculated 
as a function of a geometric parameter $\theta\in[-1,1]$ that measured particle shape, with $\theta=\pm 1$ for 
uniaxial rod and plate geometries, respectively. One-component monolayers of prolate or oblate freely-rotating 
ellipsoids were also recently studied via the Parsons-Lee DF and Molecular-Dynamics simulations \cite{szabi}. In particular, 
the effect of orientational restriction of ellipsoids on the orientational properties of monolayers 
was studied. The ground states of monolayers of hard ellipsoids interacting through a quadrupole 
pair-potential were recently found \cite{Heinemann}. Apart from the T-like configurations, three more 
particle orientations were predicted to be stable. 

Here we extend our previous model \cite{yuri1} to a binary mixture of uniaxial rods and plates. 
Board-like-shaped rods and plates are taken to be symmetric: although they have different eccentricities 
(prolate and oblate), their volumes and their aspect ratios (ratio between major and minor 
particle edge-lengths) are taken to be the same. This choice of shape geometries is motivated 
by their extensive use in studies of the biaxial-nematic (B) phases stability with respect to nematic-nematic 
(N-N) phase separation in binary \cite{roij,varga1,varga2} and polydisperse \cite{yuri2,yuri3} 
mixtures of rods and plates. We are interested in the effect of particle adsorption on the phase behaviour 
of the mixture, in an effort to elucidate (i) the propensity of the system to phase separate into 
different phases, (ii) the nature (second vs. first order) of the N-B transition when the adsorption 
strengths are changed, (iii) the relative stability of the B phase with respect to the N or other 
non-uniform phases, and (iv) the representative phase diagrams of the system (calculated for certain selected values of 
model parameters). In general we found a rich phase behaviour with the presence of 
two disconnected demixed N-B phase transitions, one located at low pressures, with a B phase rich in 
rods, and the other at very high pressures, with a B phase rich in plates (although B-B demixing 
also occurs in a narrow range of pressures). When plates are strongly adsorbed,  
the mixture exhibits a strong first-order phase transition. Different demixing scenarios depend on 
the relative values of plate and rod adsorption coefficients. When adsorption of rods is large as 
compared to that of plates, the B-phase stability is greatly enhanced, the N-B transition is always of 
second order, and no demixing occurs.   

The article is organized as follows. In Sec. \ref{model} we introduce the model and the theoretical tools 
used to perform the calculations. In Sec. \ref{results} we summarize all the results obtained, with
subsections presenting different mixtures with various relative adsorption strengths and particle aspect ratios. 
Finally some conclusions are drawn in Sec. \ref{conclusions}.  

\section{Model}
\label{model}
\begin{figure}
\epsfig{file=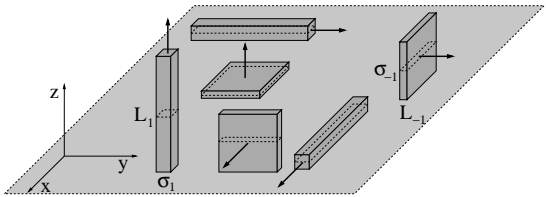,width=2.8in}
\caption{Schematic representation of the Zwanzig model of a symmetric rod-plate mixture adsorbed on a monolayer.}
\label{fig1}
\end{figure}

We use the Zwanzig model for a binary mixture of prolate (rods) and oblate (plates) board-like particles 
with centres of mass lying on the $xy$ plane and with main axes pointing along the $\nu=x,y,z$ directions. 
The edge lengths of species $s$ ($s=1$ for rods and $s=-1$ for plates) 
are represented through the tensor 
\begin{eqnarray}
\sigma_{s\nu}^{\tau}=\sigma_s+\left(L_s-\sigma_s\right)\delta_{\nu\tau},
\end{eqnarray}
with $\delta_{\nu\tau}$ the Kronecker delta symbol, while $L_s$ and $\sigma_s$ are the particle sizes parallel and perpendicular
to the main particle axis, respectively. Therefore, particles are uniaxial parallelepipeds with a square section of area 
$\sigma_s^2$. Only symmetric mixtures will be studied, namely those composed of particles with the same volume
(which is set to unity, $v_0=L_s\sigma_s^2=1$), and with aspect ratios $\kappa_1$ and $\kappa_{-1}$
related by $\kappa\equiv \kappa_1=L_1/\sigma_1=\kappa_{-1}^{-1}=\sigma_{-1}/L_{-1}\geq 1$. Thus the edge lengths 
of the different species are calculated as $L_s=\kappa^{2s/3}$, $\sigma_s=\kappa^{-s/3}$. See Fig. 
\ref{fig1} for a schematic representation of our model.

The theory used in the present calculations is the uniform limit of the FMT density functional obtained by applying the dimensional crossover property.
This feature allows to correctly transform the functional from 3D to 2D by assuming a 3D density profile, for species $s$ and orientation along the $\nu$-axis, 
of the form $\rho_{s\nu}^{\rm(3D)}({\bm r})\equiv \rho_{s\nu}\delta(z)$, i.e. imposing that the particle centres of mass 
are constrained to the flat surface perpendicular to $z$. When this density is substituted into the 3D version of the excess free-energy functional, the resulting 
functional depends on the (constant) 2D number densities
\begin{eqnarray}
\rho_{s\nu}=\rho x_s \gamma_{s\nu}, 
\end{eqnarray}
where $\rho$ is the total number density, $x_s$ is the molar fraction of species $s$, while  $\gamma_{s\nu}$ 
is the fraction of species $\nu$ with main axes pointing along the $\nu$-direction. Obviously these 
quantities fulfill the following constraints: $\displaystyle\sum_s x_s=1$ and $\displaystyle\sum_{\nu} \gamma_{s\nu}=1$ or alternatively
$\displaystyle\sum_s \rho_s=\rho$ and $\displaystyle\sum_{\nu} \rho_{s\nu}=\rho_s$, with $\rho_s=\rho x_s$ the number density of species $s$. 
The resulting excess free-energy density in reduced thermal units $kT$ depends on the following weighted 
densities:
\begin{eqnarray}
&&n_0=\rho=\sum_{s,\nu}\rho_{s\nu}, \quad n_2=\eta=\sum_{s,\nu}\rho_{s\nu} \sigma_{s\nu}^x\sigma_{s\nu}^y,\\
&&n_{1x}=\sum_{s,\nu}\rho_{s\nu}\sigma_{s\nu}^x,\quad n_{1y}=\sum_{s,\nu}\rho_{s\nu}\sigma_{s\nu}^y,
\end{eqnarray}
and has the explicit form 
\begin{eqnarray}
\Phi_{\rm exc}\equiv \frac{\beta{\cal F}_{\rm exc}}{A}=-n_0\log(1-n_2)+\frac{n_{1x}n_{1y}}{1-n_2},
\end{eqnarray}
where $\beta=1/kT$ is the Boltzmann factor and $A$ the total area of the system.
We note that the weighted density $n_2$ is just the total packing fraction, $\eta$, of the binary mixture. 
The ideal part, $\Phi_{\rm id}\equiv \beta{\cal F}_{\rm id}/A$ is, as usual
\begin{eqnarray}
\Phi_{\rm id}= \sum_{s,\nu} \rho_{s\nu}\left(\log \rho_{s\nu} -1\right),
\end{eqnarray}
and the effective interaction between species $s$, with projected area on the $xy$ plane 
$a_{s\nu}=\sigma^x_{s\nu}\sigma^y_{s\nu}$, and the surface, is accounted for by an external potential 
contribution to the free-energy density: 
\begin{eqnarray}
\Phi_{\rm ext}\equiv \frac{\beta {\cal F}_{\rm ext}}{A}=-\sum_{s\nu}\epsilon_s \rho_{s\nu} a_{s\nu}.
\end{eqnarray}
Note that this contribution is proportional to the projected particle areas, and that we allow for the possibility that the adsorption strengths
$\epsilon_s\geq 0$ be dependent on species $s$.  

To find the equilibrium orientational properties of the fluid we minimize the total free-energy density 
$\Phi=\Phi_{\rm id}+\Phi_{\rm exc}+\Phi_{\rm ext}$ with respect to the fractions $\gamma_{s\nu}$. These can be related to
the uniaxial nematic order parameters,
\begin{eqnarray}
Q_s\equiv \frac{3\gamma_{sz}-1}{2}, \quad s=\pm 1,
\end{eqnarray} 
which measure the order about the direction perpendicular to the surface, and to the biaxial nematic order parameters,
\begin{eqnarray}
\Delta_s=\frac{s\left(\gamma_{sx}-\gamma_{sy}\right)}{\gamma_{sx}+\gamma_{sy}}, \quad s=\pm 1,
\label{biaxial}
\end{eqnarray}
which measure the degree of biaxial order. 
Note that the factor $s=\pm 1$ in the definition of $\Delta_s$ is necessary to take account of
the orthogonality of the plate and rod main axes when their projections have the same orientations. 

We have calculated the phase diagram by searching for possible demixing transitions through (i) the equality 
between chemical potentials of species $s$, 
$\displaystyle{\beta \mu_s\equiv \frac{\partial\Phi}{\partial \rho_s}}$,  and (ii) the equality between the
pressures of the demixed phases. The latter can be calculated as 
\begin{eqnarray}
\beta p=\frac{n_0}{1-n_2}+\frac{n_{1x}n_{1y}}{(1-n_2)^2}
\end{eqnarray}
(in reduced thermal units). These calculations are equivalent to finding the coexisting molar fractions of the demixed phases through 
the double-tangent construction of the Gibbs free-energy density,
\begin{eqnarray}
\beta g(x)\equiv \Phi+\beta p\rho^{-1},
\end{eqnarray}
as a function of the molar fraction $x\equiv x_1$ at constant value of $p$. 
This constraint allows us to find the total number density $\rho(x,p)$ as a function of $x$ at fixed $p$, 
while all the quantities $\gamma_{s\nu}$ should be calculated for the same values $(x,p)$ from 
the set of equations $\partial \Phi/\partial \gamma_{s\nu}=0$. To find the location of second-order phase 
transitions to biaxial phases we use a bifurcation analysis of the total free-energy density $\Phi$ 
with respect to the (small) order parameters $\Delta_s$. For details of these calculations see Appendix 
\ref{append}. In the rest of the manuscript we use dimensionless densities $\rho^*\equiv \rho v_0^{2/3}$ and pressures 
$p^*\equiv \beta p v_0^{2/3}$.

\section{Results}
\label{results}

\begin{figure}
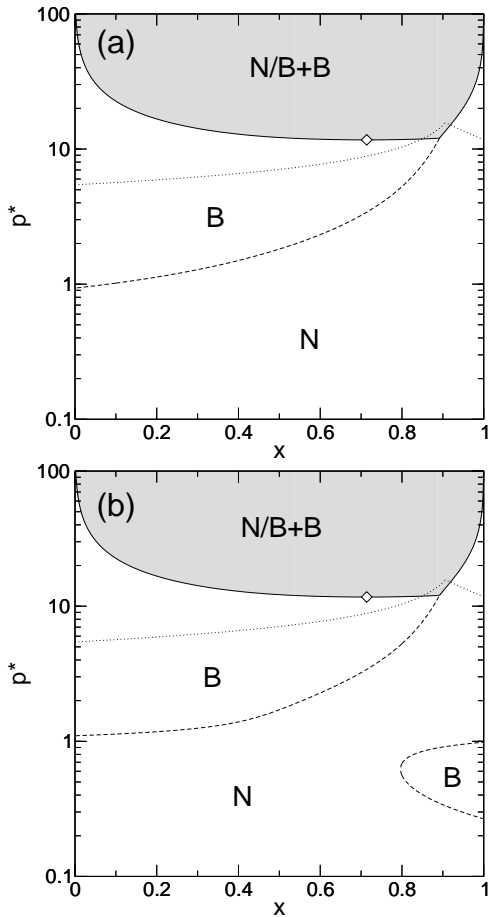

\epsfig{file=fig2a.eps,width=2.5in}
\epsfig{file=fig2b.eps,width=2.5in}
\caption{Phase diagrams of rod-plate binary mixtures with $\kappa=10$ shown in the pressure-molar fraction of 
rods ($x\equiv x_1$) plane. The values of the external potential strengths in (a) and (b) are 
$(\epsilon_1,\epsilon_{-1})=(0,0)$ and $(1,1)$, respectively. The dashed lines show the second order N-B 
transitions, while the solid lines show the binodals of N-B or B-B coexistences. The grey-shaded region 
(labelled as N/B+B) is the demixing region. The open diamond corresponds to the demixing critical point. 
The dotted lines are the spinodal instability to non-uniform phases. The regions of stability 
of the N and B phases are correspondingly labelled.}
\label{fig2}
\end{figure}

This section is devoted to the study of the phase diagram topologies as a function of 
the adsorption strengths $(\epsilon_1,\epsilon_{-1})$ and the aspect ratio $\kappa$ of the mixture.

\subsection{Mixtures with $\kappa=10$ and low and symmetric adsorption} 

We firstly studied the monolayer of rods and plates with zero adsorption to the surface by setting 
$(\epsilon_1,\epsilon_{-1})=(0,0)$. The phase diagram obtained for a mixture with $\kappa=10$ 
is shown in Fig. \ref{fig2}(a) in the pressure-composition plane. The dashed line, which departs at its 
lowest pressure from the left vertical axis ($x\equiv x_1\sim 0$), represents a continuous N-B phase transition. 

From low to high pressures, but below the N-B spinodal, the configuration of plates changes from a 
nearly equimolar composition of their three species to that in which the species with the largest (square) projected area, equal 
to $\kappa^{2/3}$, has the lowest composition while the other two species, with rectangular projected shapes of aspect ratio $\kappa$ 
and surface area $\kappa^{-1/3}$, have equal compositions. The plate axes of the latter species are parallel to the surface but their 
rectangular sections are yet randomly oriented in 2D. This phase is a planar N with a negative uniaxial 
order parameter which decreases with pressure. The configuration of rods at low pressures always exhibits
a preferential alignment perpendicular to the monolayer, resulting in a higher proportion of projected (small) squares 
of surface area $\kappa^{-2/3}$. The other two species of rods, having rectangular shapes, 
aspect ratio $\kappa$, and surface area $\kappa^{1/3}$, have again the same composition. Thus the uniaxial 
order parameter of rods is always positive and increases with pressure. The authors of the recent work
\cite{Oettel} showed that the uniaxial order parameter $Q_1$ of one-component rods on a surface increases linearly from zero as a function 
of $\rho^*$. 

\begin{figure}
\epsfig{file=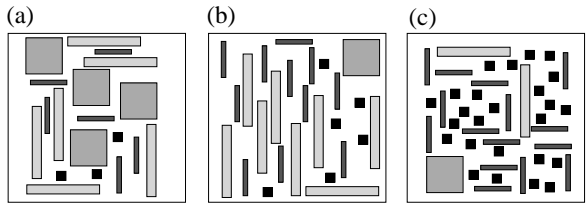,width=3.in}
\caption{Schematic representation of projected particle areas corresponding to rod-plate binary mixtures 
in (a) a low-density N phase, (b) an intermediate-density B phase, and (c) a high-density N phase.}
\label{fig0}
\end{figure}

As the pressure increases, the fraction of plates with their axes oriented parallel to the surface and that 
of rods oriented perpendicular to it become larger, and the uniaxial order parameters tend to $-1/2$ and $1$, 
respectively. At a certain pressure (which depends on molar fraction), the $xy$ orientational symmetry 
is broken in a continuous fashion, and the projected rectangular species for both, rods and plates, begin to 
align along a preferential direction, say the $y$ direction. From this pressure the B phase becomes stable and  
the B order parameters continuously increase from zero. See Fig. \ref{fig0} for a sketch of projected shape 
configurations for the different phases. Fig. \ref{fig2} indicates that biaxial ordering is promoted by the plates: 
the N-B spinodal, departing from the left vertical axis ($x\equiv x_r=0$), is a monotonically increasing function of $x$ 
and possesses an asymptote at $x=1$. The latter means that one-component monolayers of rods do not exhibit B ordering. 
From now on we refer to this spinodal as the {\it plate} N-B {\it spinodal}. 

At very high pressures, most of the plate axes align parallel to the monolayer, while those of rods are 
perpendicular to it. Thus the system can be approximated by a 2D mixture of Zwanzig hard rectangles 
(of area $\kappa^{-1/3}$ and aspect ratio $\kappa$) and parallel hard squares (of area $\kappa^{-2/3}$). 
This 2D mixture demixes into two phases, each one rich in one of the species, as shown in Fig. \ref{fig2}. 
The critical point of the demixing transition is above the N-B spinodal, meaning that B-B coexistence 
takes place in some pressure interval. For pressures above the crossing point between 
the right demixing binodal and the N-B spinodal the system phase separates into a B phase, rich in plates, 
and a N phase, rich in rods. Finally the dotted line in Fig. \ref{fig2} represents
the spinodal instability of the uniform phases with respect to density modulations, which corresponds to the 
presence of stable non-uniform phases (see Appendix \ref{append1} for details on these calculations). 
We can see that the N/B-B demixing is, except for a small 
interval of pressures, metastable with respect to transitions to non-uniform phases. 

Fig. \ref{fig2} (b) shows the phase diagram when the adsorption strengths are still relatively small: 
$(\epsilon_1,\epsilon_{-1})=(1,1)$. We can observe that the phase diagram topology is similar to that of the 
preceding case, except that now there appears a region, close to $x=1$ and bounded by a dashed line, where the B phase 
becomes stable. From now on this spinodal will be called the {\it rod} N-B {\it spinodal}.
The total free-energy is lowered when the fraction of rods with main axes parallel the monolayer increases, since
this is proportional to the projected areas. In turn, these rods exhibit two continuous N-B and B-N transitions: 
B ordering increases with pressure, reaches a maximum, then decreases and finally disappears altogether. 
This reentrant behaviour of the B phase with pressure can be explained as follows.
When the loss in free-energy given by a preferential adsorption and further alignment of rods with projected rectangular shapes 
cannot compensate the free-energy increase due to the large excluded volumes between 
rectangular projected species, as compared to those of small squared species, the most favored configuration 
of particles is that of rods pointing perpendicular to the monolayer. As the total amount of rods lying on the surface becomes 
small a B-N transition takes place. This behaviour was already found in monolayers of one-component 
Zwanzig rods with $\kappa>\kappa_c$ and zero adsorption. The values of $\kappa_c$ were found to be 21.3 
and 12 from the spatially continuous \cite{yuri1} and discrete lattice models \cite{Oettel}. 
When the continuous orientational degrees of freedom are restored, this transition disappears \cite{szabi}. 
However if particles can rotate freely except for a small solid angle with respect to the surface normal, 
the B phase again becomes stable \cite{szabi}. Thus in real situations when the surface/interface 
promotes a preferential adsorption of rods with their axes parallel to the surface the B phase will 
certainly become stable. 

\begin{figure}
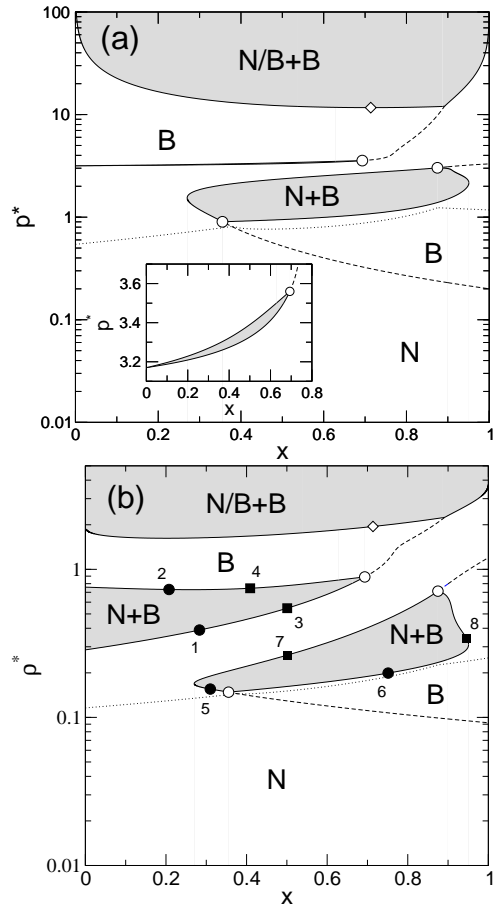

\epsfig{file=fig4a.eps,width=2.5in}
\epsfig{file=fig4b.eps,width=2.5in}
\caption{Phase diagrams of rod-plate binary mixtures for $\kappa=10$ and $(\epsilon_1,\epsilon_{-1})=(4,4)$ 
in the (a) pressure-molar fraction plane, (b) scaled density-molar fraction plane. Solid lines show 
the binodals of N-B (or B/B) coexistence, with the coexistence regions shaded in grey. The dashed, dotted 
lines and diamond have the same meaning as in Fig. \ref{fig2}. Open circles correspond to tricritical 
points, while black squares and circles in (b) are used to show a pair of coexisting points at different 
binodals. Inset in (a) is a zoom showing the first-order character of the N-B transition.}
\label{fig3}
\end{figure}

\begin{figure}
\epsfig{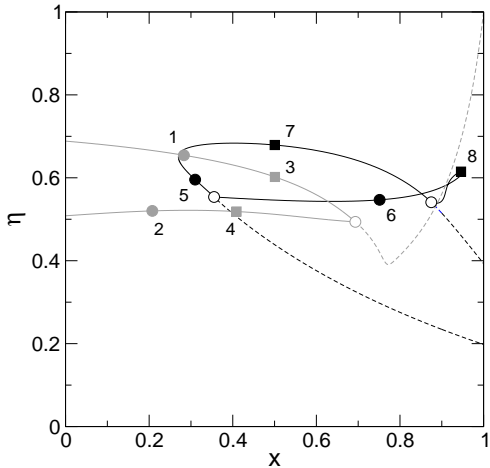}
\caption{Packing fractions corresponding to the N-B coexistence binodals (solid) and N-B second-order 
transitions (dashed) of a rod-plate binary mixture with $\kappa=10$ and $(\epsilon_1,\epsilon_{-1})=(4,4)$ 
as a function of molar fraction. The symbols have the same meanings as in Fig. \ref{fig3} and correspond 
to the same state points. Packing fractions corresponding to the lines belonging to the 
low pressure N-B demixing and to the strong first order N-B transition of 
Fig. \ref{fig3}(b) are here shown in grey and black, respectively.}
\label{fig4}
\end{figure}

\subsection{Mixtures with $\kappa=10$ and high and symmetric adsorption}

Now we proceed to describing the phase behaviour of monolayers of rods and plates with relatively 
high adsorption strengths, specifically those with $(\epsilon_1,\epsilon_{-1})=(4,4)$. The phase diagram in 
the pressure-composition and total density-composition planes are shown in Fig. \ref{fig3}(a) and (b), 
respectively. The most salient features that can be observed from the figure are: (i) the presence of 
strong N-B demixing at pressures located between two tricritical points, both lying on the rod-N-B  
spinodal which ends at $x=1$, and (ii) the existence of a strong first order N-B transition departing 
from $x=0$ and ending in a tricritical point located at the plate-N-B spinodal. Panel (b) shows the density 
of the coexisting phases along all these binodals and spinodals. 

\begin{figure}
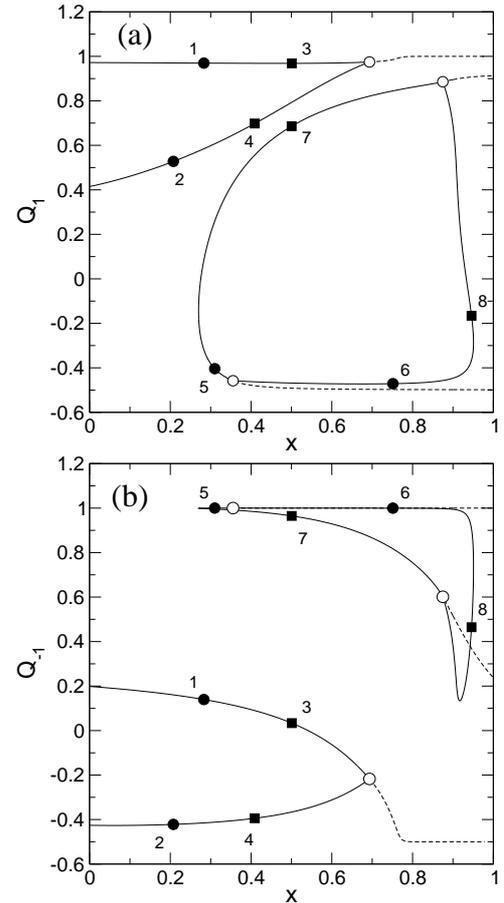

\epsfig{file=fig6a.eps,width=2.5in}
\epsfig{file=fig6b.eps,width=2.5in}
\caption{Uniaxial orientational order parameters of (a) rods, and (b) plates as a function of the molar 
fraction along the N-B coexisting binodals and spinodals of Fig. \ref{fig3}, i.e. for a rod-plate binary 
mixture with $\kappa=10$ and $(\epsilon_1,\epsilon_{-1})=(4,4)$. The solid and dashed lines, and 
symbols (showing the same state points) have the same meanings as in Fig. \ref{fig3}.}
\label{fig5}
\end{figure}

\begin{figure}
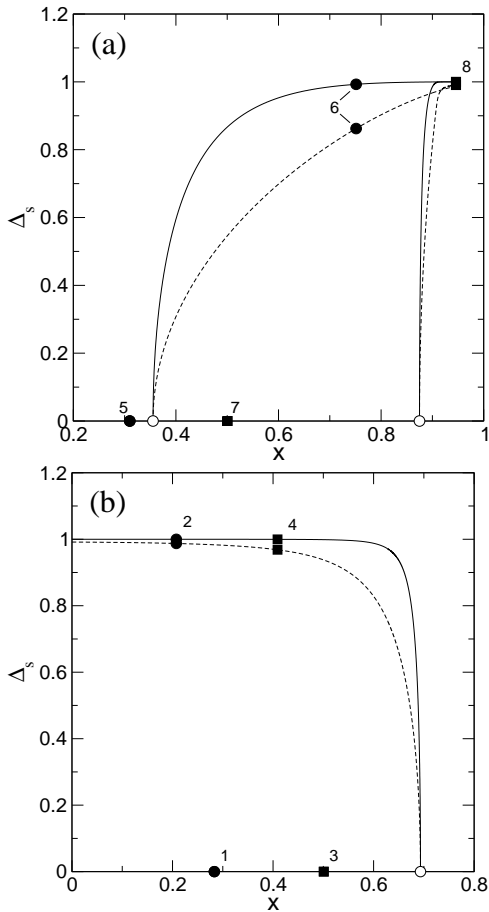

\epsfig{file=fig7a.eps,width=2.5in}
\epsfig{file=fig7b.eps,width=2.5in}
\caption{Biaxial orientational order parameter of rods (solid) and plates (dashed) 
as a function of the molar fraction along the N-B demixing binodals (a) and along the N-B strong first order 
transition (b) corresponding to  Fig. \ref{fig3}, i.e. for the rod-plate binary mixture with 
$\kappa=10$ and $(\epsilon_1,\epsilon_{-1})=(4,4)$. The symbols have the same meanings as in Fig. 
\ref{fig3} and represent the same state points.} 
\label{fig6}
\end{figure}

We first describe the N-B demixing. Fig. \ref{fig3}(b) shows that the values   
of the coexisting densities in the demixed phases are similar, with B being the densest phase,  
rich in rods (two different pairs of coexisting densities are shown with circles and squares). 
However the packing fraction has the opposite behaviour (see the black line of 
Fig. \ref{fig4}): the phase with the highest packing fraction is the N phase, rich in plates. This behaviour 
(nearly the same coexisting densities but very different composition) is typical in entropy-driven 
demixing. Note that plates in both coexisting N and B phases have always a positive uniaxial order 
parameter along the binodals [see Fig. \ref{fig5}(b)], so the fraction of projected large squares (cross-section of plates) 
is relatively high. The projected rectangles corresponding to rods lying 
on the surface also have a relatively high fraction (see the negative values of $Q_1$ in (a) along the 
binodals, except for a region close to the upper tricritical point), as compared to that of the 
square projected areas (when rods point perpendicular to the monolayer).
As usually occurs in entropy-driven demixing, the total excluded area between the rod-projected 
rectangles and the plate-projected squares is lowered if the demixed phases are rich in one of the species. 
The high proportion of large squares is the reason behind the high packing fraction values of the coexisting N phase,
as compared to that of the B phase. It is interesting to note the highly non-monotonic behaviour of the packing fraction 
along the demixing binodals (see Fig. \ref{fig4}), a direct consequence of the dependence of $\eta$ not 
only on $\rho^*$ and $x$ but also on the order parameters $Q_s$. Finally Fig. \ref{fig6} shows the 
evolution of the biaxial order parameters along the demixing binodals. The biaxial 
order of both species along the coexisting B phase rapidly increases and saturates to its highest value (unity) 
as the mixture gets away from the lower tricritical point. At some point they invert their monotonicity and decrease
to zero at the upper tricritical point.

From the other side of the phase diagram ($x=0$), and for intermediate values 
of pressure, a strong first order N-B phase transition takes place, as can be inferred from the large coexisting 
density gap in Fig. \ref{fig3}(b). This transition ends in a tricritical point located at the 
plate-N-B spinodal, and is driven by the reorientation of plates. Note the positive and negative values of 
$Q_{-1}$, corresponding to the coexisting N and B phases, respectively [see  Fig. \ref{fig5}(b)]. 
The rods are now mainly oriented perpendicular to the monolayer (perfectly oriented in the N phase,
and with a small degree of orientation in the B phase). It is interesting to note that the coexisting 
molar fractions are now similar (with the B phase slightly rich in plates), while the coexisting densities 
are very dissimilar (B being the densest phase). This transition is driven by a differential change 
in free-energy from a N phase with a high fraction of adsorbed larges squares (plate axes perpendicular to the monolayer) 
to a B phase with a high fraction of projected rectangles (corresponding to plates with their axes lying on the monolayer and pointing along $y$). 
When we follow a constant pressure path from the N to the B coexisting phases, the free-energy 
contribution corresponding to the external potential increases, while that coming from the entropic 
interaction part is lowered (because the particle excluded areas decrease). The differential change in the 
total adsorbed area of particles is huge, so the transition becomes strongly first order. Again 
the coexisting packing fractions are inverted: that of the B phase is lower (see the gray solid curve in 
Fig. \ref{fig4}). At very high pressures the same N/B-B demixing transition ending in a critical point 
takes place [similar to the cases $(\epsilon_1,\epsilon_{-1})=(0,0)$ and $(1,1)$].

Finally we calculated the instability of uniform phases with respect to non-uniform density modulations. 
The pressures and densities at which these instabilities occur are plotted as a function of $x$ in 
Fig. \ref{fig3}(a) and (b), respectively. We can see that the lower tricritical point is located 
above this curve, suggesting that all demixing transitions are metastable with respect to 
transitions  or demixing between non-uniform phases. When the molar fraction of plates with axes 
perpendicular to the surface is high due to their large surface adsorption,  
their square cross-sections may crystallize in a simple square lattice at a certain pressure. 
If pressure is increased beyond this value, plates will reorient their axes parallel to the monolayer 
and consequently crystal ordering could be destabilized with respect to a uniform or nonuniform phase exhibiting B ordering. 
On the other side of the phase diagram, where the molar 
fraction of rods is high and the B phase is stable, the most likely scenario is that of B-smectic or B-columnar 
phases, where the projected rectangles exhibit two-dimensional smectic or columnar arrangements. 
These open questions should be settled out by performing DF minimisation with respect to
non-uniform density profiles, $\rho_{s\nu}(x,y)$, and search for possible coexistences, a formidable 
task that we leave for future studies. 

\subsection{Mixtures with $\kappa=10$ and asymmetric adsorption}

\begin{figure}
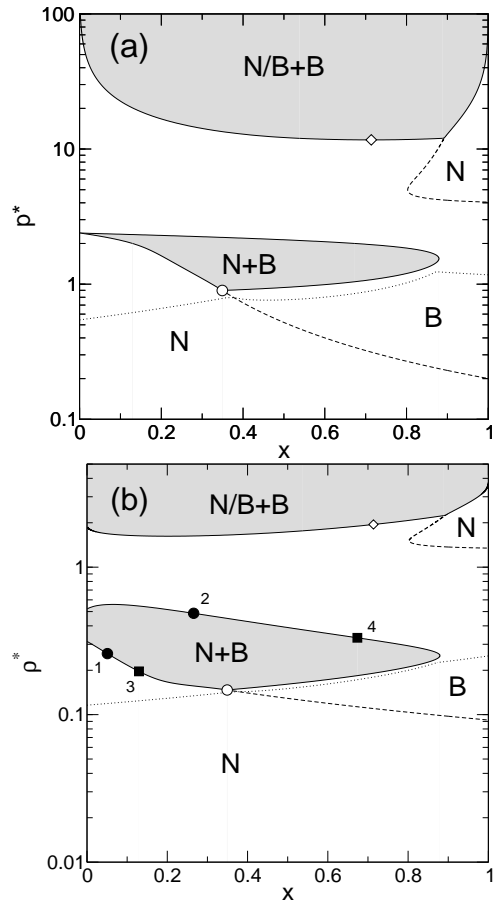

\epsfig{file=fig8a.eps,width=2.5in}
\epsfig{file=fig8b.eps,width=2.5in}
\caption{Phase diagrams in the pressure-molar fraction [(a)] and scaled density-molar fraction [(b)] 
planes for a rod-plate binary mixture with $\kappa=10$ and  $(\epsilon_1,\epsilon_{-1})=(5,3)$. The lines, 
symbols and labels have the same meanings as in Fig. \ref{fig3}.}
\label{fig7}
\end{figure}

\begin{figure}
\epsfig{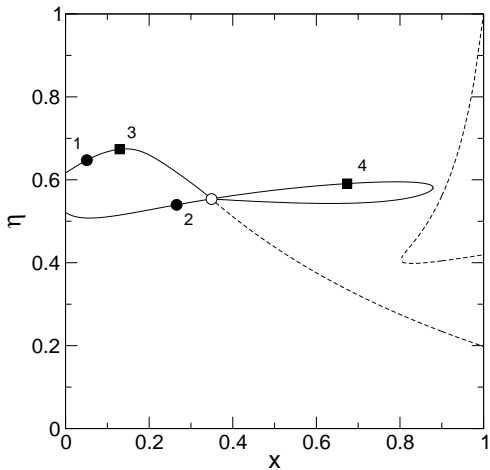}
\caption{Packing fraction, as a function of the molar fraction, along the N-B binodals and spinodals 
of Fig. \ref{fig7}, i.e. for the rod-plate mixture with $\kappa=10$ and $(\epsilon_1,\epsilon_{-1})=(5,3)$.
The lines and symbols have the same meaning as in Fig. \ref{fig7}, and they correspond to the same state 
points.}
\label{fig8}
\end{figure}

\begin{figure}
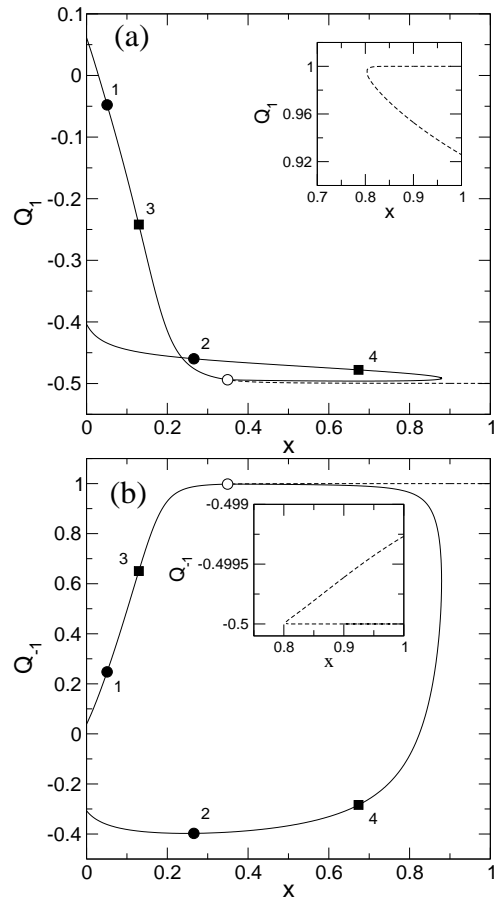

\epsfig{file=fig10a.eps,width=2.5in}
\epsfig{file=fig10b.eps,width=2.5in}
\caption{Uniaxial nematic order parameter of (a) rods, and (b) plates along the B binodals and spinodals 
of Fig. \ref{fig7}, i.e. for the rod-plate binary mixture with $\kappa=10$ and 
$(\epsilon_1,\epsilon_{-1})=(5,3)$. The lines and symbols have the same meaning as in Fig. 
\ref{fig7}. The inset in (a) shows the order parameter corresponding to the highly oriented rods along the N-B 
spinodal located at high pressures and close to $x=1$ in Fig. (\ref{fig7}). The inset in (b) shows the 
order parameter of plates with close-to-perfect planar nematic ordering along the same N-B spinodal.}
\label{fig9}
\end{figure}

\begin{figure}
\epsfig{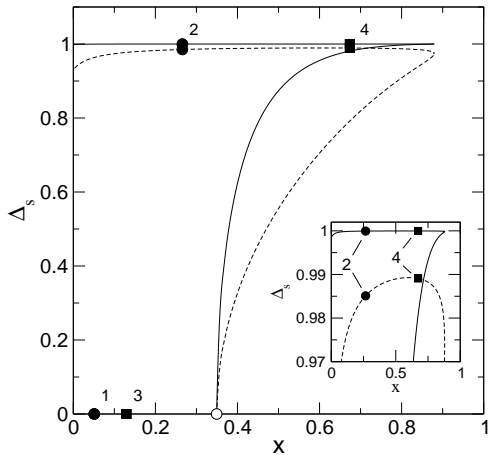}
\caption{Biaxial orientation order parameter of rods (solid curve) and plates (dashed curve) along the 
B binodals of Fig. \ref{fig7}, i.e. for a rod-plate binary mixture with $\kappa=10$ and 
$(\epsilon_1,\epsilon_{-1})=(5,3)$. The symbols have the same meaning as in Fig. \ref{fig7}, and show 
the same state points. The inset is a zoom of the main figure.}
\label{fig10}
\end{figure}

In this section we study the effect of adsorption asymmetry on the phase behaviour of monolayers 
of rods and plates. To this purpose we have chosen the adsorption strengths as 
$(\epsilon_1,\epsilon_{-1})=(5,3)$, i.e. rods are more strongly adsorbed on the surface than plates. To better compare 
the results obtained with those described in the preceding sections, we again set the aspect ratio to
$\kappa=10$. The phase diagram for this mixture is plotted in Fig. \ref{fig7} in the (a) pressure-composition 
and (b) density-composition variables. The main features we can extract from these results are: 
(i) When the rods are strongly 
adsorbed on the surface, the lower part of the rod-N-B spinodal meets the plate-N-B spinodal at intermediate 
compositions, creating a monotonic, fully connected spinodal curve over the whole composition interval. 
(ii) There is a lower tricritical point located on this curve, above which a demixing transition occurs. 
(iii) The demixing transition coalesces with the strong first-order transition driven by plates at 
higher pressures (the one ending at $x=0$). (iv) The upper part of the rod-N-B spinodal joins
the right part of the plate-N-B spinodal, creating an island of N phase stability. (v) The 
entropic N/B-B demixing at high pressure remains invariant, which confirms the fact that the system behaves like a 2D mixture of 
squares and rectangles. Fig. \ref{fig8} shows the strong non-monotonic behaviour of the packing fraction 
along the coexistence binodals, with the presence of a large loop. (vi) The spinodal for the uniform phase 
instability with respect to density modulations is located again below the tricritical point. 
Also, packing fraction inversion (with respect to density) does occur, with N being the densest phase. 
Figs. \ref{fig9}(a) and (b) show the uniaxial order parameters along the binodals, with a B phase 
of rods and plates having axes lying on the monolayer, and a N phase with plate axes pointing perpendicular to it. 
Interestingly, the rods in N phase are oriented parallel to the surface, although to a lesser degree. 
The insets show the completely saturated ordering of particles (perpendicular and 
parallel to the surface for rods and plates, respectively) along the boundaries limiting the island 
of N phase stability. Finally Fig. \ref{fig10} shows the biaxial order parameters along the coexisting 
binodals, which have the usual behaviour: a rapid increase as the system gets away from the tricritical point, and then 
saturation to perfect biaxial ordering. 

\begin{figure}
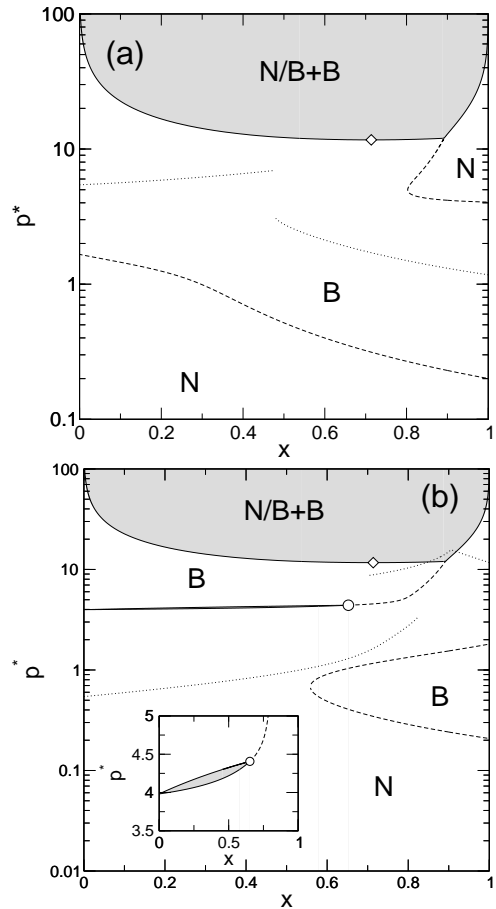

\epsfig{file=fig12a.eps,width=2.5in}
\epsfig{file=fig12b.eps,width=2.5in}
\caption{Phase diagrams in the pressure-molar fraction planes corresponding to
rod-plate binary mixtures with (a) $\kappa=10$ and $(\epsilon_1,\epsilon_{-1})=(5,1)$, and (b)
$(2,5)$. The lines, symbols and labels have the same meaning as in Fig. \ref{fig3}. Inset in (b) is a zoom 
of the main figure showing the first-order character of the N-B transition in some pressure interval.}
\label{fig11}
\end{figure}

Fig. \ref{fig11}(a) shows the phase diagram of a mixture with $\kappa=10$, but with larger asymmetry 
in their adsorption strengths, $(\epsilon_1,\epsilon_{-1})=(5,1)$. Now the plates are slightly adsorbed 
on the surface, while rods are strongly adsorbed. We can see that: (i) the demixing and first-order N-B transitions 
(present in the $(5,3)$-mixtures) are substituted by continuous transitions, 
with the N-B spinodal now being a monotonically decreasing function of $x$. Thus the region of B stability 
is greatly enhanced. (ii) The island of N-phase stability and the N/B-B demixing at high pressure
remain as before. Fig. \ref{fig11}(b) shows the phase 
diagram of the mixture $(\epsilon_1,\epsilon_{-1})=(2,5)$, i.e. plates and rods are strongly and 
slightly adsorbed, respectively. Note the presence of a strong first-order N-B transition driven by the 
desorption of plates at intermediate pressures. As before, this transition ends in a tricritical point located on 
the plate-N-B spinodal. The B phase, rich in rods, is again stable inside a island bounded by the 
rod-N-B spinodal. No demixing was found in this part of the phase diagram, with the N-B transition 
being of second order. The N/B-non-uniform-phase spinodals are now discontinuous and located above 
the B phase of rods. 

\subsection{Mixtures with $\kappa=$20 and 40}

\begin{figure}
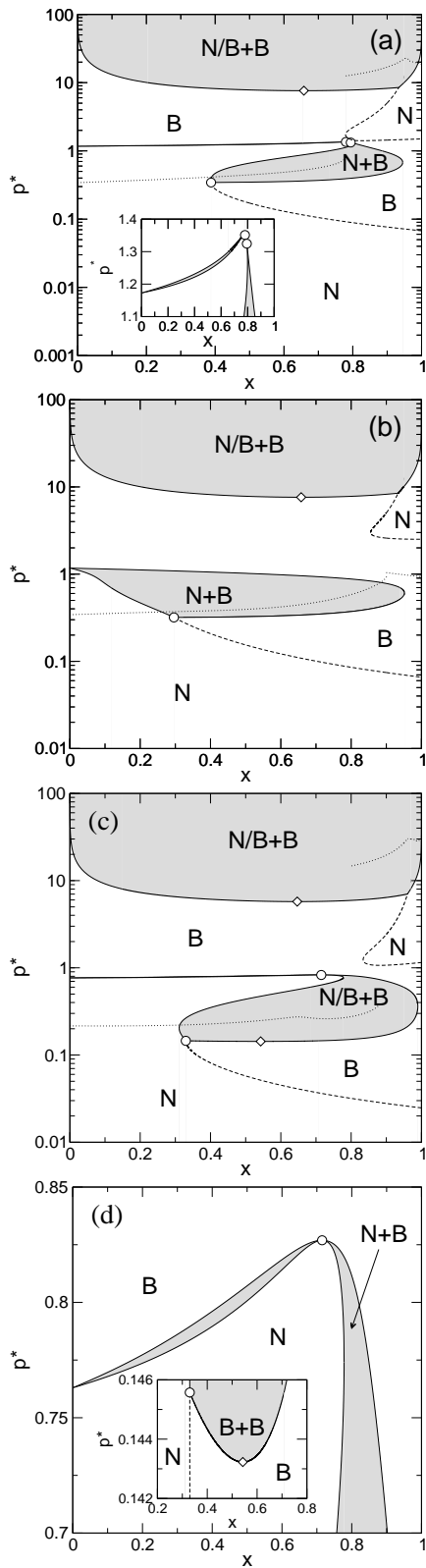

\epsfig{file=fig13a.eps,width=2.2in}
\epsfig{file=fig13b.eps,width=2.2in}
\epsfig{file=fig13c.eps,width=2.2in}
\epsfig{file=fig13d.eps,width=2.2in}
\caption{Phase diagrams in the pressure-molar fraction plane of rod-plate binary 
mixtures for (a) $\kappa=20$, $(\epsilon_1,\epsilon_{-1})=(1.5,1.5)$, (b) $\kappa=20$, $(\epsilon_1,\epsilon_{-1})=(3,1.5)$,
(c) and (d) $\kappa=40$, $(\epsilon_1,\epsilon_{-1})=(1,1)$.
The lines, symbols and labels have the same meaning as in Fig. \ref{fig3} and 
\ref{fig7}. In panel (d) we show two enlarged regions of the phase diagram shown in (c): one close to
the N-B azeotropic point (main figure), and the other close to the lower B-B critical point (inset).}
\label{fig12}
\end{figure}

The last study concerns the phase behaviour of mixtures with higher aspect ratios, in particular those 
with $\kappa=20$ and 40. As shown in Fig. \ref{fig12} the phase-diagram topologies are similar, but 
there is an important difference: now the lower tricritical point is always located below the spinodal 
instability to non-uniform phases. Thus there is always a range of pressures, which increases with $\kappa$, 
for which demixing into a N phase and a B phase rich in rods is stable. 
In Figs. \ref{fig12} (a) and (b) the phase diagrams for $\kappa=20$ and 
$(\epsilon_1,\epsilon_{-1})=(1.5,1.5)$, panel (a), and $(3,1.5)$, panel (b), are shown. Note that the former mixture
is symmetric with respect to adsorption.
The upper boundary of the rod-N-B spinodal meets the N-B-plate spinodal at intermediate compositions. The phase 
behaviour includes: N-B demixing between two tricritical points (the lower one departing from 
the rod-N-B spinodal), and a strongly first-order N-B transition driven by the desorption of plates, 
beginning at $x=0$ and ending in a tricritical point located at the plate-N-B 
spinodal. This point and the upper critical point of the demixing transition are now very close to each other. 
An island of N-phase stability exists at high pressure as a consequence of the coalescence 
between the right part of the plate-N-B spinodal and the upper part of the rod-N-B spinodal.  
Finally, N/B-B demixing of the effective two-dimensional mixture of squares and 
rectangles at high pressure is also present. The phase diagram topology for the second mixture studied, that with 
$(\epsilon_1,\epsilon_{-1})=(3,1.5)$, is very similar to the for $\kappa=10$
described before. Again there is an important difference, namely the existence of stable N-B demixing 
in some pressure range. Also the B phase, rich in rods, is stable in a rather large region of 
the phase diagram.

We end this section by showing the phase diagram of a mixture 
with $\kappa=40$ and $(\epsilon_1,\epsilon_{-1})=(1,1)$, in Figs. \ref{fig12}(c) and (d). 
We concentrate on some details of the phase diagram topology not found before: (i) The presence of a 
lower critical point (at low pressures and close to the lower N-B tricritical point) above which B-B 
demixing takes place in a rather small range of pressures [see inset of panel (d)]. (ii) The two distinct
tricritical points, located close to each other in previous cases [e.g. the case $\kappa=20$ 
and $(\epsilon_1,\epsilon_{-1})=(1.5,1.5)$] now coalesce into a single azeotropic point [see panel (d)]. 

\section{Conclusions}
\label{conclusions}

We have systematically studied the phase behaviour of mixtures of rods and plates adsorbed on a monolayer. In our model, 
the particle centres of mass are taken to freely move on the surface, while particles can  
rotate in 3D within the restricted-orientation, Zwanzig approximation. Adsorption of the particle surfaces on the monolayer 
is mimicked through an attractive external potential proportional to the area of the particle surface contact, while the strengths of this 
interaction, $(\epsilon_1,\epsilon_{-1})$, depend on the species type. Rods and plates were taken to be symmetric, 
i.e. with the same volume and same aspect ratio $\kappa$. A FMT-based DF, adapted to the present constrained geometry, was minimised, and
phase diagrams of mixtures with $\kappa=10$, 20 and 40 and different values of $(\epsilon_1,\epsilon_{-1})$ were calculated.

The main results can be  summarized as follows. (i) When both adsorption strengths $\epsilon_s$ are zero or small,
rods and plates orient perpendicular and parallel to the surface, respectively. 
The degree of orientation continuously increases with pressure, and at some value (which depends on the composition) a second-order 
N-B transition occurs, at which the plate axes orient along a director lying on the surface. 
Although rods also exhibit biaxial ordering, this transition is governed by the orientational 
symmetry breaking of plate axes, and consequently the plate-N-B spinodal is a monotonically increasing function of $x$.
(ii) At some value of $\epsilon_1$ and, starting at the $x=1$ vertical axis at low pressures, there appears an island of 
B phase stability enclosed by a rod-N-B spinodal which is disconnected 
from the plate-N-B spinodal. The N-B transition is of second order and is governed by the alignment 
of rods with axes on the monolayer. As pressure increases rods prefer 
to align perpendicular to the surface, and biaxial ordering disappears. 
(iii) When rods and plates are symmetrically adsorbed on the monolayer ($\epsilon_1=\epsilon_{-1}$) 
and the strengths are relatively large, two tricritical points appear on the rod-N-B spinodal; between these two points
N-B demixing takes place, with demixed phases rich in one of the components. 
The B and N coexisting phases are mostly populated by rods and plates, respectively. Also, there is a strongly first-order 
N-B phase transition with a large density gap starting at the $x=0$ vertical axis. This 
transition ends in a tricritical point located on the plate-N-B spinodal, and is driven 
by the desorption of the largest cross-section of the plates (corresponding to axes perpendicular to the monolayer). 
Finally at very high pressures, when the degree of order is high, the system effectively becomes a 
two-dimensional mixture of squares (the smallest projected section of rods) and rectangles (the smallest projected
sections of plates), which demix into a B phase, rich in plates, and a N phase, rich in rods. This demixing 
transition ends in a critical point above which there exists a rather narrow B-B demixing region.
(iv) When the adsorption of particles is very asymmetric and  $\epsilon_1>\epsilon_{-1}$, the lowest 
boundary of the rod-N-B spinodal connects with the left part of the plate-N-B spinodal, forming 
a monotonically-decreasing spinodal over the whole range of compositions. The N-B transition is always of 
second order. Also the upper boundary of the rod-N-B spinodal connects with the right part of the plate-N-B-plate 
spinodal, forming an island of N stability. The highest pressure N/B-B demixing remains invariant.
(v) When the adsorption of particles is very asymmetric and $\epsilon_{-1}>\epsilon_1$, the strongly
first-order N-B transition governed by desorption of plates and their B ordering is present 
up to high molar fractions, while the N-B transition, governed by orientation of rods, is of second order 
and the N-B demixing disappear. The region of stability of the B phase (enclosed by the rod-N-B spinodal) 
is reduced as $\epsilon_1$ becomes smaller. 

We also calculated the spinodal instability of uniform phases with respect to density modulations with
different symmetries (smectic, columnar or crystalline). We found that a B phase rich in rods is  
stable over a relatively large interval of pressures, while the strong N-B phase transition is always metastable. 
For $\kappa=20$ and 40 there exists a range of pressures for which  N-B demixing is stable.  
Of course N/B-B demixing at high pressure, ending in a lower critical point, is also metastable. We note that
demixing regions could be made wider if we chose shape-asymmetric mixtures (different volumes and/or 
different aspect ratios), and consequently the regions of N and B phase stability could be modified. 
Even the orientational symmetries of the demixed phases could be different for asymmetric mixtures as  
shown in theoretical calculations of freely-rotating rod-plate Onsager mixtures \cite{Wensink}.

We are confident that the results presented in this work will be qualitatively similar if we remove 
the restricted orientation approximation and consider the free rotation of particle axes. 
Computer simulations of binary mixtures adsorbed on a flat monolayer could confirm this conclusion.  

MC simulations of 2D mixtures of rods on a lattice show an interesting phase behavior \cite{Kundu}. When the 
aspect ratio 
of the longer rods is 7 there exist two I-N and N-I transitions  as the density of longer rods is increased 
while that of the shorter rods is fixed bellow a certain critical density. This behavior resembles that of the present 
rod-plate mixture for which two N-B and B-N transitions take place at fixed composition as the pressure is increased
and the adsorption strengths of rods is high enough. Thus it would be interesting to perform DFT calculations on 
mixtures of adsorbed rods to find the differences and similarities between monolayers and strictly 2D hard rod mixtures.

Also interesting are the similarities that  the present system shares with the phase
behaviour of monolayers of biaxial particles studied in \cite{miguel1}. In that case
the phase diagrams, in the density-biaxial parameter plane, present a N-B spinodal, completely
analogous to that of the present system, if we replace the biaxial parameter by 
molar fraction. Moreover, by increasing the aspect ratio of biaxial rods in the previous study, one obtains
an island of B phase stability; a similar effect is found in the present study when adsorption strength 
is increased. Above a particular value for the largest aspect ratio of the biaxial particles, this island coalesces
with the N-B spinodal, again a behaviour similar to the one found in the present study if the
adsorption strength is increased beyond some critical value. 

Additionally, we may ask ourselves how particle biaxiality would affect the present phase
behaviour. If particles are biaxial, we would expect the N-B spinodal to shift to higher
pressures, favouring the stabilisation of non-uniform phases. On the other hand, biaxiality
would also reduce the island of B stability in the rod-rich part of the phase diagram.
Finally it would also be possible that particle biaxiality reduced demixing gaps,
because particle projections become similar with biaxiality.

\appendix

\section{N-B bifurcation analysis}
\label{append}

The constrained minimisation of the free-energy density with respect to the variables $\gamma_{s\nu}$
gives the following set of equations:
\begin{eqnarray}
\gamma_{s\nu}=\frac{e^{-\Psi_{s\nu}}}{\displaystyle\sum_{\tau} e^{-\Psi_{s\tau}}},
\label{itera}
\end{eqnarray}
where 
\begin{eqnarray}
&&\Psi_{sx}=\left(\frac{n_0}{1-n_2}+\frac{n_{1x}n_{1y}}{(1-n_2)^2}-\epsilon_s\right)\kappa^{s/3}\nonumber\\
&&+\frac{n_{1x}\kappa^{-s/3}+
n_{1y}\kappa^{2s/3}}{1-n_2},\label{uno}\\
&&\Psi_{sy}=\left(\frac{n_0}{1-n_2}+\frac{n_{1x}n_{1y}}{(1-n_2)^2}-\epsilon_s\right)\kappa^{s/3}\nonumber\\
&&+\frac{n_{1x}\kappa^{2s/3}+
n_{1y}\kappa^{-s/3}}{1-n_2},\\
&&\Psi_{sz}=\left(\frac{n_0}{1-n_2}+\frac{n_{1x}n_{1y}}{(1-n_2)^2}-\epsilon_s\right) \kappa^{-2s/3}\nonumber\\
&&+\frac{(n_{1x}+n_{1y})}{1-n_2}\kappa^{-s/3}.\label{tres}
\end{eqnarray}
Using the definition of biaxial order parameter $\Delta_s$ from (\ref{biaxial}), we arrive at
\begin{eqnarray}
\Delta_s=\frac{e^{-\Psi_{sx}}-e^{-\Psi_{sy}}}{e^{-\Psi_{sx}}+e^{-\Psi_{sy}}},\quad s=\pm 1.
\label{itera1}
\end{eqnarray}
Expanding (\ref{itera1}) with respect to $\Delta_s$ (obviously the functions $\Psi_{s\nu}$ depend on 
$\{Q_s,\Delta_s\}$) up to first order gives a system of 
two equations which has a nontrivial solution only if the number density is such that
\begin{eqnarray}
\rho^{-1}=\frac{1}{2}\sum_s x_s \kappa^{-2s/3}\left[\left(\kappa^{2s}+1\right)+\gamma_{sz}
\left(1-\kappa^{2s}\right)\right],
\end{eqnarray}
where $\gamma_{sz}$ is the solution of (\ref{itera}) for $\nu=z$, and the functions $\Psi_{s\nu}$ 
are calculated from (\ref{uno})-(\ref{tres}), with all the weighted densities depending only on 
$\gamma_{sz}$, as $n_0=\rho$ and
\begin{eqnarray}
&&n_{1x}=n_{1y}\nonumber\\
&&=\frac{\rho}{2}\sum_s x_s\left[\kappa^{2s/3}+\kappa^{-s/3}+\gamma_{sz}\left(
\kappa^{-s/3}-\kappa^{2s/3}\right)\right],\nonumber\\&&\\
&&n_2=\eta=\rho\sum_s x_s\left[\kappa^{s/3}+\gamma_{sz}\left(\kappa^{-2s/3}-\kappa^{s/3}\right)\right] 
\end{eqnarray}
Thus we need to solve a system of two non-linear equations to find the equilibrium values of $\gamma_{sz}$,
and consequently the number density $\rho$ at which a bifurcation from the N to the B phase occurs.

\section{Spinodal instability to non-uniform phases}
\label{append1} 

We have calculated the spinodal instability of uniform phases with respect to spatial inhomogeneities 
through the divergence of the structure factor. The Fourier transform of the direct correlation functions, 
calculated through the second functional derivative of the functional, reads
\begin{eqnarray}
&&-\hat{c}_{s\mu,s'\nu}({\bm q})=
\frac{\langle \hat{\omega}^{(0)}_{s\mu}({\bm q})
\hat{\omega}^{(2)}_{s'\nu}({\bm q})
+\hat{\omega}^{(1x)}_{s\mu}({\bm q})\hat{\omega}^{(1y)}_{s'\nu}({\bm q})\rangle}
{1-n_2}\nonumber\\
&&+\frac{\langle \left(\hat{\omega}^{(1x)}_{s\mu}({\bm q})n_{1y}+
\hat{\omega}^{(1y)}_{s\mu}({\bm q})n_{1x}\right)\hat{\omega}^{(2)}_{s'\nu}({\bm q})\rangle}
{(1-n_2)^2}\nonumber\\
&&+\left(\frac{n_0}{(1-n_2)^2}+\frac{2n_{1x}n_{1y}}{(1-n_2)^3}
\right)\hat{\omega}^{(2)}_{s\mu}({\bm q})\hat{\omega}^{(2)}_{s'\nu}({\bm q}),
\end{eqnarray}
where $\langle f_{s\mu,s'\nu}\rangle=f_{s\mu,s'\nu}+f_{s'\nu,s\mu}$ implies symmetrisation with respect to the 
pair of indexes $(s\mu,s'\nu)$, and the weighted densities $n_{\alpha}$ correspond to the stable uniform N or B 
phases, with orientational order parameters calculated from the minimisation of the corresponding 
free energies. The Fourier transforms of the weighting functions are
\begin{eqnarray}
&&\hat{\omega}^{(0)}_{s\mu}({\bm q})=\chi_0(q_x\sigma^x_{s\mu}/2)\chi_0(q_y\sigma^y_{s\mu}/2),\\ 
&&\hat{\omega}^{(2)}_{s\mu}({\bm q})=\sigma_{s\mu}^x\sigma_{s\mu}^y\chi_1(q_x\sigma^x_{s\mu}/2)\chi_1(q_y\sigma^y_{s\mu}/2),\\
&&\hat{\omega}^{(1x)}_{s\mu}({\bm q})=\sigma_{s\mu}^x\chi_1(q_x\sigma^x_{s\mu}/2)\chi_0(q_y\sigma^y_{s\mu}/2),\\
&&\hat{\omega}^{(1y)}_{s\mu}({\bm q})=\sigma_{s\mu}^y\chi_0(q_x\sigma^x_{s\mu}/2)\chi_1(q_y\sigma^y_{s\mu}/2),
\end{eqnarray}
where $\chi_0(x)=\cos x$ and $\chi_1(x)=\sin(x)/x$. We define the $6\times 6$ structure factor matrix 
\begin{eqnarray}
S^{-1}_{\alpha,\beta}({\bm q})=\delta_{ss'}\delta_{\mu\nu}-\rho_{s\mu}\hat{c}_{s\mu,s'\nu}({\bm q}),
\end{eqnarray}
where $\alpha=s+\mu+1$ and $\beta=s'+\nu+1$ if $s=-1$, and $\alpha=s+\mu+2$ and 
$\beta=s'+\nu+2$ if $s=1$ (with the corresponding relabelling $\mu=1,2$ and $3$ for $x$, $y$ and $z$, respectively). Evaluating the determinant of this matrix 
${\cal S}(q;\rho^*)\equiv \text{det}\left[S^{-1}_{\alpha,\beta}(q)\right]$ at the wave vectors 
${\bm q}=(0,q)$ or ${\bm q}=(q,0)$ (corresponding to inhomogeneities along or perpendicular to the nematic 
director, respectively), we found the corresponding values at the spinodal instabilities, $q_s$ and $\rho^*_s$, as the values for $q$ and 
$\rho^*$ where the absolute minimum of ${\cal S}(q;\rho^*)$ as a function of 
$q$ becomes zero for the first time. This is equivalent to solving the pair of equations   
\begin{eqnarray}
{\cal S}(q;\rho^*)=0,\quad \frac{\partial {\cal S}}{\partial q}(q;\rho^*)=0.
\end{eqnarray}

\acknowledgments
We acknowledge financial support from Grants FIS2013-47350-C5-1-R and FIS2015-66523-P from Ministerio 
de Econom\'{\i}a y Competitividad of Spain.

\end{document}